\documentclass[final,1p,times]{elsarticle}

 \usepackage{graphics}
 \usepackage{color}

\usepackage{amssymb}

\journal{Journal of Membrane Science}

\begin{document}

\begin{frontmatter}



\title{Water in a Polymeric Electrolyte Membrane: Sorption / Desorption and Freezing phenomena}


\author[LIPhy]{Marie Plazanet\corref{cor1}}
\ead{marie.plazanet@ujf-grenoble.fr}
\address[LIPhy]{Laboratoire Interdisciplinaire de Physique, Universite Joseph Fourier et CNRS- UMR 5588, BP87, 38402 Saint Martin d'Heres Cx, France}
\cortext[cor1]{Corresponding author. 140 rue de la physique, BP 87, 38403 Saint Martin d'h\`eres Cedex 09, France. Tel.: +33 476 51 43 35. Fax. +33 476 63 54 95}

\author[UniPG]{Francesco Sacchetti}
\address[UniPG]{Dipartimento di Fisica, Universit\`a degli Studi di Perugia, I-06123 Perugia, Italy}

\author[UniPG]{Caterina Petrillo}

\author[ILL]{Bruno Dem\'e}
\address[ILL]{Institut Laue langevin, BP 87, 38042 Grenoble Cx 9, France}

\author[LENS]{Paolo Bartolini}
\address[LENS]{European Laboratory for Non-Linear Spectroscopy (LENS), Via N. Carrara 1, I-50019 Sesto Fiorentino, Firenze, Italy}

\author[LENS,UniFz]{Renato Torre}
\address[UniFz]{Dip. di Fisica ed Astronomia, Universit\`a di Firenze, Via N. Carrara 1, I-50019 Sesto Fiorentino, Firenze, Italy}

\begin{abstract}
Nafion\textsuperscript{\textregistered} is a perfluorosulfonated polymer, widely used in Proton Exchange Membrane Fuel Cells. This polymer adopts a complex structural organisation resulting from the microsegregation between hydrophobic backbones and hydrophilic sulfonic acid groups. Upon hydration appear water-filled channels and cavities, in which are released the acidic protons to form a solution of hydronium ions in water embedded in the polymer matrix. Below 273 K, a phenomenon of water sorption/desorption occurs, whose origin is still an open question. Performing neutron diffraction, we monitored the quantity of ice formed during the sorption/desorption as a function of temperature down to 180 K. Upon cooling, we observe that ice forms outside of the membrane and crystallises in the hexagonal Ih form. Simultaneously, the membrane shrinks and dehydrate, leading to an increase of the hydronium ions concentration inside the matrix. Reversibly, the ice melts and the membran!
 e re-hydrate upon heating. A model of solution, whose freezing point varies with the hydronium concentration, is proposed to calculate the quantity of ice formed as a function of temperature. The quantitative agreement between the model and experimental data explains the smooth and reversible behavior observed during the sorption or desorption of water, pointing out the origin of the phenomena. The proposed picture reconciles both confinement and entropic effects. Other examples of water filled electrolyte nano-structures are eventually discussed, in the context of clarifying the conditions for water transport at low temperature.  

\end{abstract}

\end{frontmatter}

\begin{figure}
\includegraphics[width=12 cm]{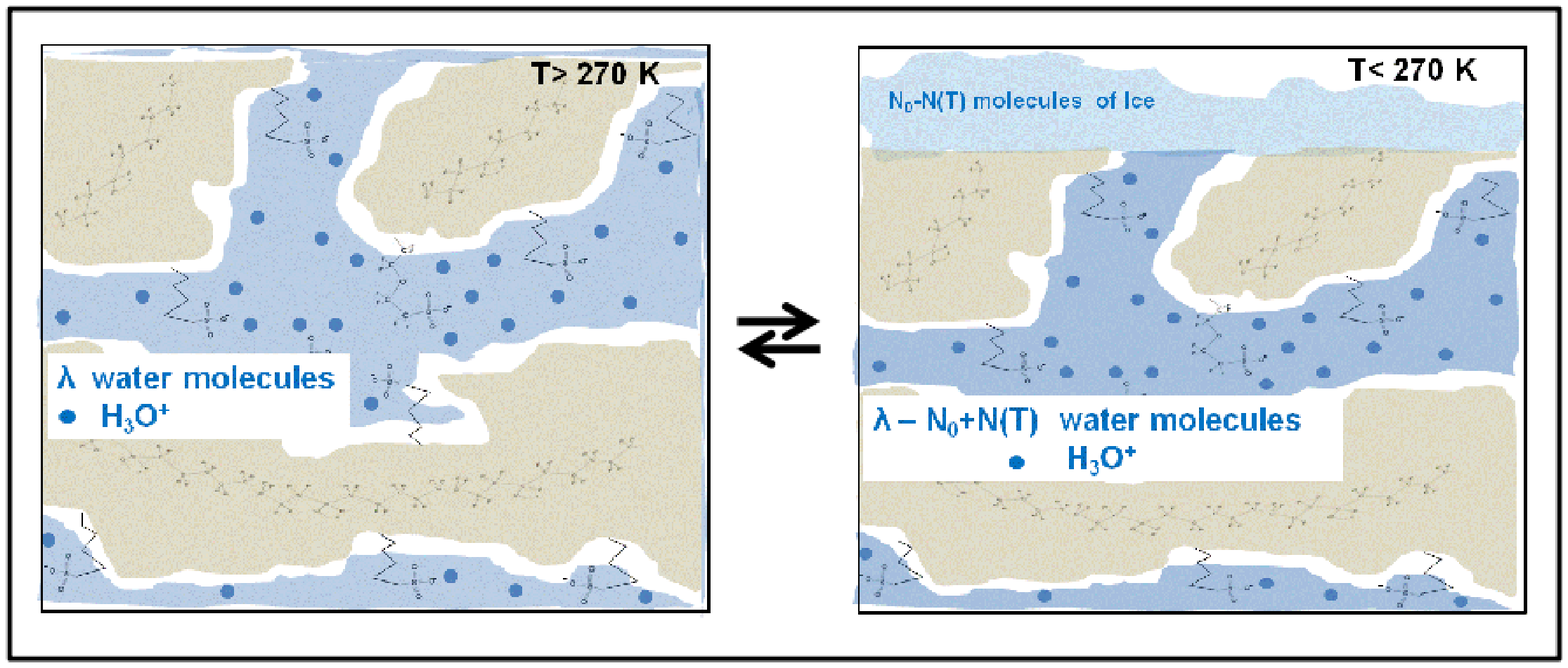}
\end{figure}



\section{Introduction}
Nafion\textsuperscript{\textregistered} is a well known amphiphilic polymer made of fluorinated hydrophobic backbone on which are distributed perfluoro vinyl ether pendant groups, terminating with hydrophilic sulfonic acid ($SO^- _3 $) anions.
It is widely used in a variety of applications \cite{Heitner1996}, the best known being the use as proton conducting membranes in fuel cells \cite{Dupuis2011}. The properties of the material are mostly related to the water structure and dynamics, that are the result of a complex polymer organization: the backbone chains tend to aggregate and form, when hydrated, a network of water-filled channels and cavities wrapped by the hydrophilic acid groups. A large amount of work has therefore been carried in view of elucidating all the properties of the material in any conditions, including variation of temperatures, hydration, acidity conditions or more. 

The structure of the material is still a matter of debate and studies. Different structural models have been proposed\cite{Mauritz2004}, from the original cluster-network model \cite{Gierke1981,Hsu1983}, to the rod-like structure\cite{Gebel2000,Rollet2002}, or a structure of ordered cylindrical nanochannels \cite{Schmidt2008}. The common feature to all of these models is the presence of water-filled spaces (i.e. cavities and/or pores) inside a polymeric matrix wrapped by the sulfonate groups. The characteristic  dimensions of these cavities or pores depend on the amount of adsorbed water, typically spanning from 2 nm to about 6 nm. The water adsorbed by the polymer matrix can be defined by the “hydration number” $\lambda$ = number of water molecules per sulfonate group. The lambda  value that depends on several chemical-physical parameters \cite{Maldonado2012,Choi2005II,Cappadonia1994,Cappadonia1995} as temperature and pressure; and by some material characteristics: as e!
 quivalent weight (i.e. moles of sulfonate groups per grams of dry membrane) and heat treatment of the membrane.

The water dynamics taking place inside the polymeric electrolyte membrane (PEM) is also an important issue that to date remains open, various diffusion mechanisms are proposed for the water and hydronium ions \cite{Choi2005,Paciaroni2006,Perrin2007,Feng2011}. The water dynamics is particularly relevant for the conductivity capabilities that is one of the key features for PEM applications. These have been investigated in Nafion membrane under various conditions, in particular at subzero temperatures where both temperature and hydration influence the conductance \cite{Cappadonia1994, Cappadonia1995}. According to these results, water inside PEMs, and in particular in Nafion, remains mobile below 273 K suggesting that it enters into a supercooled liquid state. The mobility can persists at much lower temperatures dependent on PEM characteristics and hydration level. 

A further question in PEM studies is the characterization and understanding of water sorption/desorption mechanism and water crystallization appearing at subzero temperatures. The dehydration of the membrane upon cooling below 273 K was indeed first highlighted in the eighties \cite{Pineri1984, Pineri1985}, followed later by several more detailed studies \cite {Yoshida1992,Corti2006,Thompson2006,Pineri2007,Plazanet2009,Guillermo2009,Gebel2011,Teocoli2012}. 
In particular the one performed by Pineri et al., see ref. \cite{Pineri2007}, enabled to characterise the water desorption mechanism and ice formation, and measure the time scales over which these processes take place. They showed that water maintains a mobility that enables the system to reach an equilibrium through desorption and ice formation on the membrane surface down to 220 K. The timescale for desorption kinetics are slow, typically at 220 K the desorption process could take more than one hour in Nafion\textsuperscript{\textregistered}117 membranes of 175 $\mu$ thickness. For a hydration number of the membrane up to  $\lambda \sim 40$, the water desorbs and forms ice outside of the pores \cite{Pineri2007,Plazanet2009}. When the membrane reaches a very high hydration level ($\lambda \sim 50$), the water eventually crystallizes inside the membrane \cite{Gebel2011}. The crystallisation, in this last case, was interpreted in terms of freezing point depression with respec!
 t to bulk water due to the water confinement, accordingly to the Gibbs-Thomson equation of a liquid confined in a pore. The possibility of a freezing point depression due to the acidic solution, i.e. hydronium ions in water, was considered but ruled out in this case. Indeed, a complex interplay between the confinement, acidity, and sorption/desorption mechanisms has to be elucidated in the PEM. Lowering the temperature the water is desorbed, this reduces the nano-cavity dimensions and increase the water acidity inside the PEM; both these effects produce a freezing point depression. The effect of confinement on the water freezing point was also proposed to explain a transition in the conductance below 273 K \cite{Cappadonia1994, Cappadonia1995}.

Our previous study \cite{Plazanet2009} was based on the propagation of high frequency acoustic waves in a highly hydrated membrane ($\lambda \sim 30$) using optical Transient Grating (TG) spectroscopy. The detection of an acoustic wave propagating at the velocity of sound in ice was used to characterize the ice formation down to $\sim 220 K$. The low acoustic damping proved the formation of micrometer size crystals, that can grow only outside of the membrane, on the surface. The amount of external ice follows a surprisingly smooth and continuous behaviour as a function of temperature, both on cooling and heating. The amount of grown ice could be phenomenologically modelled by a power law of the difference $(T-T_c)$, $T_c$ being the pseudo-critical temperature below which the ice phase exists. This smooth and reversible behaviour indicates that the phenomenon, in our case, cannot not be simply related to the freezing of confined water, but is indeed related to a more complex !
 mechanism involving the desorption/sorption processes.
 
The present study is dedicated to the understanding of the origin of the water desorption mechanisms at sub-zero temperature. Following the description of the materials and methods used for this work, we will present the Neutron Diffraction results as a function of temperature, and argue for a fitting of the data based on an ideal solution model, and compare with a confinement effect. In the last part, we will discuss the generality of this behaviour in charged systems containing confined water.

\section{Materials and Methods}
{\bf Sample preparation.}

Nafion 112\textsuperscript{\textregistered} membrane of 54 $\mu$ thickness was purchased from Ion Power Inc. (USA). The sample was cleaned according to the same procedure described in details in reference \cite{Plazanet2009}. After cleaning, the membranes were rinsed and boiled several times in $D_2O$. Careful weighting of the sample enabled the determination of $\lambda \sim 40$. The membranes were then cooled to liquid nitrogen temperature and ground, in order to get a powder of millimeter size grains and to ensure good orientational averaging in neutron diffraction experiment. They were then placed in a closed quartz cuvette under inert atmosphere avoiding any contamination by air moisture.

{\bf Neutron scattering experiment.} Neutron diffraction was performed in the Q-range of $0.02 - 2.0 $  \AA$ ^{-1}$ on the small momentum transfer diffractometer D16 (ILL, Grenoble, France), using a wavelength of 4.75 \AA$\,$  and an angular range of [$2-125^o$]. The sample was placed in a cryostat with quartz windows and the temperature was controlled to within 0.1 K. $D_2O$ hydrated samples were used in order to decrease the incoherent background arising from hydrogen atoms and to increase the coherent scattering at the origin of the diffraction pattern.

The sorption / desorption kinetics are known to be temperature dependent and slow, although much faster in membranes directly in contact with liquid water than in membrane equilibrated with water vapour \cite{Majsztrik2007, Kusoglu2012a}. The necessary waiting time for the sample to reach equilibrium in similar conditions were previously determined by Transient Grating spectroscopy\cite{Plazanet2009}. Identical waiting time has been used in the present experiments.

The diffraction measurements also enabled to confirm that no excess water was present in the sample, by checking that no Bragg peaks assigned to ice were present at 277.15 K, the melting point of $D_2O$.

\section{Experimental Results}
\begin{figure}
\includegraphics[width=12cm]{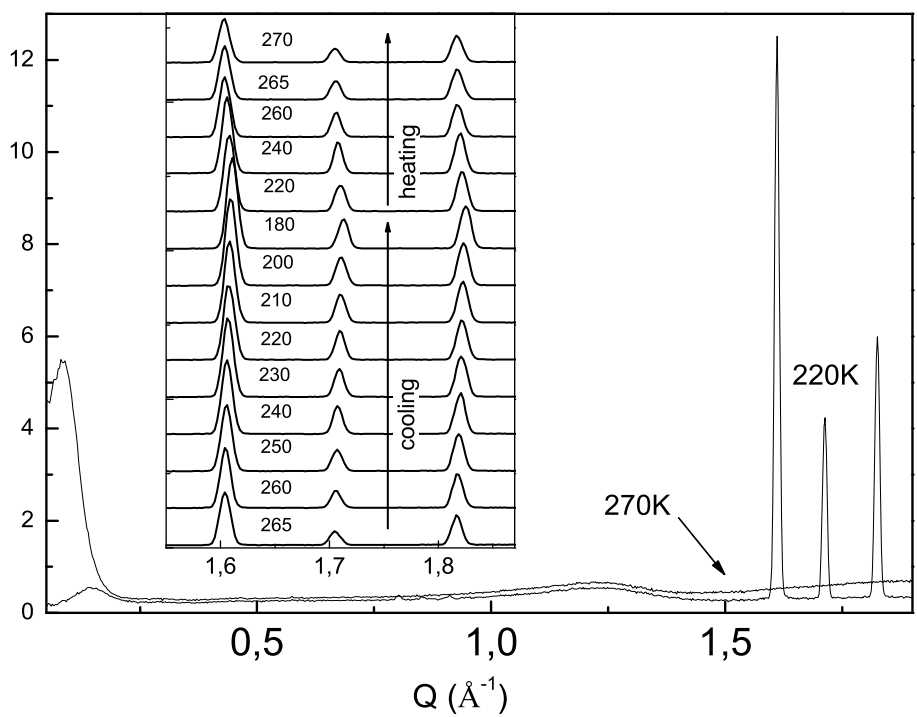}
\caption{Diffraction patterns measured on D16 (ILL) at 270 K (upon cooling, i.e. before ice formation) and 220 K. The inset shows the three peaks characteristics of hexagonal ice as a function of temperature upon cooling and heating. Each pattern is shifted by the same quantity for clarity.} \label{fig_diff}
\end{figure}

Neutron Diffraction investigations using $D_2O$ hydrated membranes enabled to quantitatively monitor the ice formation over an extended temperature range, providing information on the structural features of the ice crystallites and the Nafion membrane. The sample with a water content of $\lambda_0 \sim 40$ was prepared according to the procedure reported in the ``Material and Methods'' section and put in a tight container. The measurements were performed on complete temperature cycles, with waiting times longer than 20 minutes between measurement points separated by 10 degrees, in order to reach equilibrium. Diffraction patterns measured during cooling and heating are shown in Figure \ref{fig_diff} and exhibit changes in the peak intensities down to $\sim $180 K. The Bragg peak intensities in the region of Q $\geq$ 1.5 \AA$^{-1}$ are directly related to the amount of ice, varying with temperature. This range of temperatures over which water mobility is observed is indeed in !
 agreement with measurements of mechanical properties \cite{Teocoli2012} that show an elastic energy dissipation due to unfrozen water down to 180 K. 
At any temperature, the widths of the Bragg peaks are limited by the instrumental resolution. A rough estimation of the crystallite size can be made on the basis of the Scherrer's equation \cite{Patterson1939}, which inversely relates the broadening of the Bragg peaks to the size of the crystallites:  narrow  Bragg peaks correspond to large crystallites. The equation indicates that ice crystals should be larger than 10 nm to generate Bragg peaks as narrow as the instrumental resolution. This observation is indeed in agreement with the estimation made in the previous studies, that the ice crystals have micrometric size.

The diffraction patterns also show that, within the experimental resolution, ice crystallises in the  hexagonal form (Ih). The diffraction results confirm and extend previous observations from optical TG experiments \cite{Plazanet2009}. Only an overall temperature shift in ice formation of a few degrees is observed in Neutrons experiment due to the use of $D_2O$ while optical experiments were performed on $H_2O$-hydrated samples. The inset of Figure \ref{fig_diff} represents the temperature evolution of the intensities of the three Bragg peaks characteristic of ice Ih at momentum transfer of about 1.6 \AA$^{-1}$ (100), 1.7 \AA$^{-1}$ (002) and 1.8 \AA$^{-1}$ (101). 

\begin{figure}
\includegraphics[width=13.5cm]{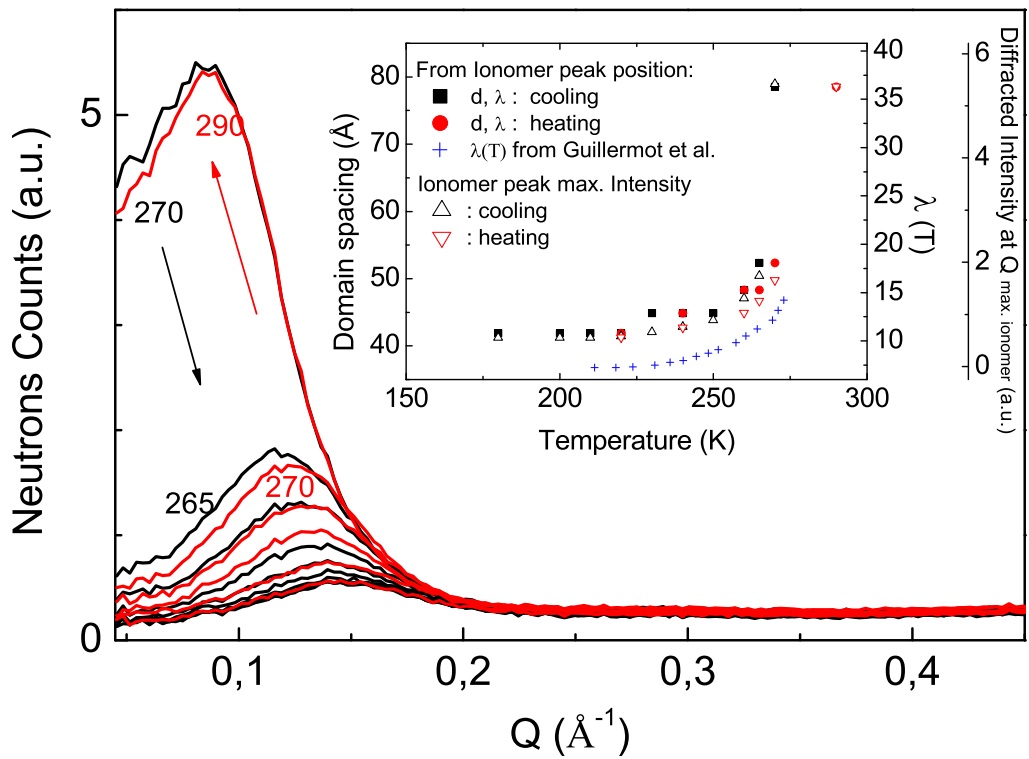}
\caption{Small angle part of the diffraction pattern, showing the ionomer peak region as a function of temperature. The inset shows the variation of the water domain spacing D during the temperature cycle calculated from the position of the peak maximum ($d=2 \pi / Q^*$), and the corresponding hydration $\lambda(T)$ estimated from the data published in the references \cite{Gebel2000, Kusoglu2012, Sood2012}. The behaviour measured by Guillermo et al.\cite{Guillermo2009} is reported for comparison. The empty symbols describe the variation of the peak intensity with temperature and are linked to the second-right scale.} \label{ionomer_peak}
\end{figure}

Beside the Bragg peaks arising from ice formation, the sorption/desorption of water is also reflected in the position of the so-called matrix ionomer peak at $ \sim 0.12 $ \AA$^{-1}$ as shown in Figure \ref{ionomer_peak}. The peak arises from the domains formed by the aggregating ionic groups, the size of the domains increasing when water swells the membrane \cite{Gebel2005}. The position of this peak, related to the distance between domains D, therefore depends on the membrane hydration, and is well documented \cite{Gebel2000, Kusoglu2012, Sood2012}. The position shifts toward higher Q upon cooling, indicating a decrease of the domains spacing and size, and follows here as a function of temperature the same trend as it does upon dehydration at room temperature, as previously shown by Guillermo et al.\cite{Guillermo2009}. This membrane contraction, as revealed by the shrinkage of the water domains, confirms that water desorbs out of the membrane and excludes the ice formatio!
 n inside the polymer matrix. Upon heating, the peak shifts back to its original position, showing that water reintegrates the membrane reversibly with temperature. It is therefore possible to have a direct evaluation of the hydration number as a function of temperature $\lambda(T)$, as shown in the inset of the Figure \ref{ionomer_peak}: the distance indeed depends, in a good approximation, linearly on the hydration number $\lambda$. A good linear fit gives the relation $\lambda=6.86d-17.8$ with d in nm. The hydration number extracted from the diffraction data at room temperature, $\lambda_0 \sim 37$, is in good agreement with weightings of the sample, that gave a number of 40 water molecules per sulfonate group. The small difference between weighting and diffraction can be easily explained by the presence of a minute amount of water on the membrane surfaces. The hydration reaches a minimal value of $\lambda_{min} \sim 11$ at 200 K, while in the data reported by Guillermo e!
 t al. it reaches a minimal value of $\lambda \sim 7.2$, independently on the initial concentration. Beside this constant offset of about $\sim 3.8$ molecules, the hydration follows however the same trend as the one reported by Guillermo et al. 

As shown by Kim et al. \cite{Kim2006}, the intensity of the peak is proportional to the difference in scattering density between the polymer matrix and the water, therefore its variation relates the variation of water content inside the membrane. This analysis holds also in our study, as is shown in the inset of figure \ref{ionomer_peak}. The intensity of the peak at its maximum position indeed follows exactly the same trend as the position, confirming that the water flows in and out of the membrane.

A small hysteresis is observed by neutron scattering in ice formation, confirming our previous results \cite{Plazanet2009}. The ice appears upon cooling only in the diffraction pattern measured at 265 K ($\sim $10 degrees below the melting point of bulk ice), and upon heating it still persists at about 270 K. This hysteresis phenomena is clearly reflected by the temperature dependence of the ionomer peak position: the data at 270 K shows a difference in the peak position of about 0.025 \AA$^{-1}$, depending if the membrane reaches this temperature upon heating or cooling, see Figure \ref{ionomer_peak}.  At 270 K, upon cooling the hydration is $\lambda \sim$ 37, i.e. the cavities are still fully hydrated and no ice is formed on the surfaces; upon heating, at the same temperature, it indicates a hydration $\lambda \sim$ 22, in other word the ice remains on the membrane surfaces reducing the amount of water inside the cavities. The hysteresis can be assigned to the supercooling!
  of the solution.

Because of the electrolyte nature of the polymer, the liquid trapped in the cavities is a solution of hydronium ions in water. At subzero temperatures, the small amount of water present on the surface can be the source of ice nucleation. Because of the percolating network of water channels in the polymer, the external ice crystallites are in contact with the liquid fractions of hydronium solution remaining inside the polymer matrix. Ice phase outside and aqueous solution inside the membrane have distinguished chemical potentials. The difference between these chemical potentials is temperature dependent, and provides the energetic contribution driving the water in and out of the membrane, which is, in other terms, an osmotic pressure.  At the freezing point of the solution, the most favourable process is to expel water out of the membrane, increasing the concentration of the hydronium ions inside the membrane and lowering further the freezing point. The water, outside, crysta!
 llises therefore following a first order transition spread over a wide temperature range. As it will be further discussed, the sorption/desorption behaviour cannot be driven by a confinement effect neither described according to the Gibbs-Thomson equation. We therefore looked closer at entropic effects. In a model of ideal solution, the melting/freezing point depression is related to the solute fraction $x_s$ (mol solute /(mol of solute +mol of water)) by the equation \cite{Laidler1995}:

\begin{equation}
ln(1-x_s) =\frac{\Delta H_{fus}}{R} \cdot \left( \frac{1}{T_0} - \frac{1}{T }\right)
\end{equation}

where R is the gas constant, $T_0$ is the bulk melting point of the pure liquid and $\Delta H_{fus}$ the enthalpy of fusion of ice. $T_0 =277.15 $ K for $D_2O$, and the temperature dependence of $\Delta H_{fus}$ is taken from reference \cite{Zavitsas2010}, and shifted by 292 J/mol for $D_2O$. From this relation can be computed the minimal concentration that avoids freezing of the solution at temperature T, and the corresponding quantity of water that is expelled out of the membrane to crystallise on the surface.

The diffracted intensity $I$ is proportional to the number of crystallised water molecules, and is normalised to 1 at the lowest temperature. $I$ can therefore be written as: 
\begin{equation}
I(T) =   \frac{1}{N_0} [  N_0- N(T) ] 
\end{equation}
where $N(T)$ is the number of water molecules in the solution at T computed from (1), and varies between 0 and $N_0$. $N(T)$ is related to the solute fraction $x_s(T)$ as follows: $x_s(T)=1/(1+N(T))$. However, $N_0$ has to be carefully chosen, as not all the water molecules present in the membrane may participate to the solution. Water properties can indeed vary between two extreme states: from water strongly bound to the surface through hydrogen bonding, to nearly free water unaffected by the interaction with the (far enough) interface, therefore classified as bulk-like water. Several classes of water might be distinguished, typically between 2 and 4 water populations depending on the experimental technique used to probe the state of water \cite{Pissis2013, Corkhill1987, Taschin2013}, on the interaction with the polymer surface \cite{Hatakeyama2010, Pineri1984} and the hydration shell of the hydronium ions \cite{Zavitsas2010}. In our sample, the maximum hydration is $\lambda_0 = 37$ and the fraction of non-freezing water molecules equals $\lambda_{min}=11$.  We therefore consider $N_0=\lambda_0 - \lambda_{min}=26$ and $N(T)=\lambda(T)-11$. 

\begin{figure}
\includegraphics[width=12cm]{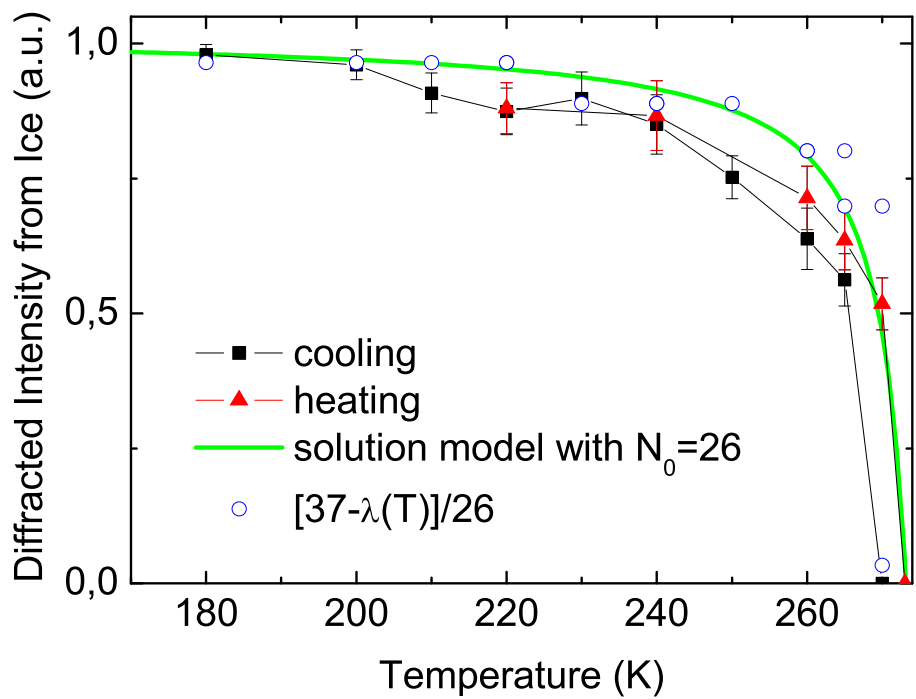}
\caption{Temperature dependence of the integrated intensity of the three diffraction peaks at 1.6, 1.7 and 1.8 \AA $^{-1}$, normalised to 1 at the lowest temperature (black full square and red full triangle correspond to intensity measurements upon cooling and heating, respectively). Solid line represents the intensity of an ideal solution in which the total number of crystallising water molecules equals 26. The empty circles are calculated using the values of $\lambda (T)$ extracted from the ionomer peak, see. figure \ref{ionomer_peak}.} \label{fig_solutionmodel}.
\end{figure}

Figure \ref{fig_solutionmodel} represents the average of the integrated intensities of the three diffraction peaks as a function of temperature, in quantitative agreement with the intensities calculated as above described. On the same plot are also reported the values of the diffracted intensity computed using $\lambda(T)$ determined from the position of the ionomer peak, as previously described. A quantitative agreement is also obtained between the model and the evaluation of the hydration state of the membrane from the ionomer peak position. 

\begin{figure}
\includegraphics[width=12cm]{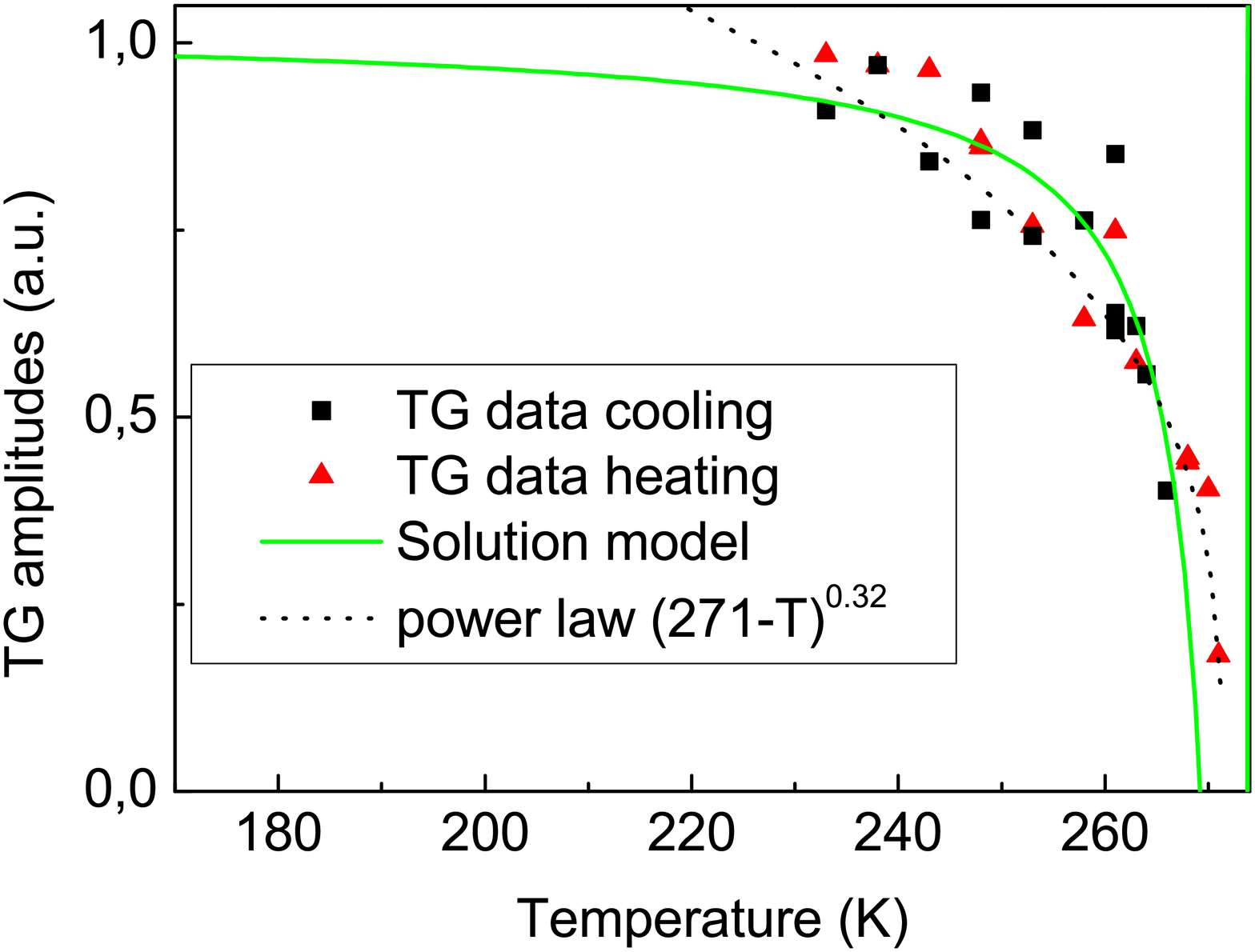}
\caption{Amplitudes of the TG data reported in Ref. \cite{Plazanet2009}. As in Figure \ref{fig_solutionmodel}, the solid line represents the amplitude from an ideal solution with $N_0= 26$. The black dotted line represent the fit with the phenomenological power law previously proposed.} \label{fig_TGdata}.
\end{figure}

The confrontation of the solution model proposed in this work, containing {\it no} fitting parameters, with the previous Transient Grating data reported in reference \cite{Plazanet2009} is presented in Figure \ref{fig_TGdata}, showing a very good agreement also with the TG data. Indeed, the TG amplitudes are directly proportional to the amount of ice formed outside the membrane and its temperature dependence can therefore be described by the solution model. The fit of the power law, previously proposed, is also reported for comparison between the two models, and is compatible with the data only in a restricted range of temperatures.

\section{Discussion}

\subsection{Confinement versus solution effect.}
Confinement effect was proposed in several studies to explain the membrane behaviour \cite{Cappadonia1994,Cappadonia1995,Gebel2011,Sivashinsky1981}. We therefore compared the freezing point depression that would be given by a pure confinement effect, i.e. according to the Gibbs-Thomson equation, with the one of the solution. The Gibbs Thomson equation predicts a freezing point that depends on the pore diameter and the surface tension between the water and the pore surface.  This concentration is directly connected to membrane hydration levels and the domain spacing D, as previously described. Moreover the confinement dimension d, assimilated to the pore/cavity diameters is related to the domain spacing D in taking into account the domain spacing of a dry membrane D$_0$=2.7 nm, by the relation d$\sim$D-2.7 \citep{Kusoglu2012,Gebel2011}.
From eq. (1), the solute concentration can be expressed as a function of $\lambda$ from $x_s=1/(1+\lambda)$, and $\lambda$ as a function of D, and d, as previously proposed. These relations enable to extract the expression of the freezing point depression as a function of pore diameter.

\begin{figure}
\includegraphics[width=12cm]{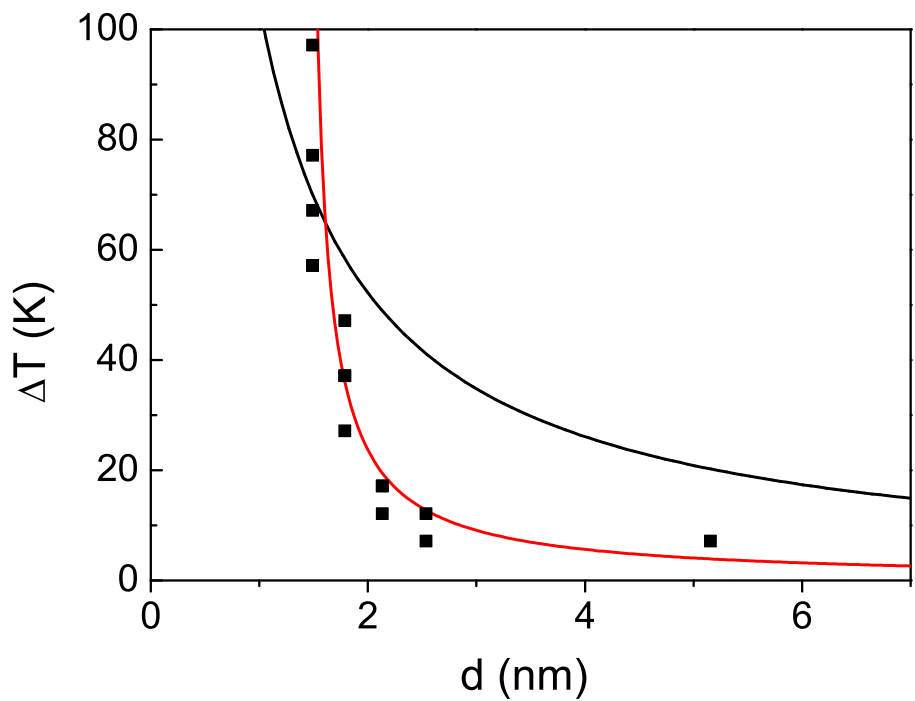}
\caption{Depression of the melting point of water as a function of the pore diameter:
(black) confined in pores of diameter d (Gibbs-Thomson eq. with adequate parameters taken from Mendel-Jakani et al.\citep{Gebel2011})
(red) in a solution of concentration varying with the domain spacing assimilated to the pore diameter. Note that $\lambda$ is shifted by 11 (and the corresponding d) because of the molecules that do not crystallise.
The model obviously fits the experimental data (reported as squares) the same way it does in the Figure \ref{fig_solutionmodel}.} \label{fig_confinement-vs-solution}.
\end{figure}

In the Figure \ref{fig_confinement-vs-solution} are compared the freezing point depression as a function of the pore diameters/domain spacing in both cases, together with our experimental points (Temperature vs. Domain Spacing, from Fig. \ref{ionomer_peak}). It appears that all the experimental data points are located below the confinement limit, meaning that the water absorbed in the membrane is always liquid because of the confinement effect. Moreover, the data are located on the freezing point curve defined by the solution model, eq (1), showing that the last effect is eventually the one driving the sorption/desorption process. Indeed, although the confinement effect is active with the thermal equilibration of the system, i.e. on a very fast timescale, the equilibration with respect to the freezing point of the solution requires sorption/desorption, which is a slow process. We therefore propose the following picture of the different processes relevant for the Nafion membr!
 ane equilibration: starting with the system in equilibrium at a given temperature $T_1$, we reduce the temperature to $T_2<T_1$. The system thermalises quickly but it is a non-equilibrium situation for the freezing of the solution, that needs to equilibrate the water amount inside/outside the membrane, on a much longer time scale. If $T_2$ is below the Gibbs-Thomson line, the confined water is supercooled and the preserved mobility enables the desorption process, driving the system toward a different hydration level and pore/cavity diameter, in agreement with the eq (1). Below $\sim$ 210 K, the confinement effect crosses the solution line, showing that confinement does not keep the water liquid any more, leading to a freezing of water inside the membrane, in agreement with all the observations of the loss of water mobility around this temperature.

We therefore propose that the main factor determining the hydration of the membrane at subzero temperature is the amount of water that can be absorbed inside the membrane according the freezing point depression line of a solution.

\subsection{Context of other systems containing confined water.}
Nafion is however not the only material for which hydration is a crucial point and where water flows at sub-zero temperatures. Hydration and dehydration of soft-condensed systems are complex processes that greatly influence their macroscopic properties. Below 273 K, the water may crystallise inside the matrix, and the ice formation can eventually lead to dehydration or damages. If crystallisation is prevented, it reaches an ultra viscous state and, if driven by some forces like in the desorption case, it can flow at temperatures of several tens of degrees below 273 K, enabling water transport at sub-zero temperature. This phenomena can be very important in many applications based on hydrogels or systems containing confined water, such as organisms cryopreservation, ion conducting membranes \cite{Neburchilov2007} or food and drug industry \cite{Zohuriaan2009}. The conditions to have reversible water desorption below 273 K in charged systems containing confined water are however not well established and deserve a particular interest.  

Water sorption and desorption has been observed in several systems as discussed below, but Nafion seems to be the first polymeric system in which low temperature water transport has been observed. Based on the phenomena observed in the Nafion material, H. Mendil-Jakani et al. checked the behaviour of water-saturated sulfonated polyimide membranes. Despite electrolyte properties similar to those of the Nafion polymer, no water desorption was observed. They therefore assigned this fact to the strong interaction between water and polymer groups, maintaining the water inside the membrane \cite{Gebel2011}. In other polymers like biocompatible ones, where the structure and properties of water have also been extensively studied as they are believed to be a major factor for biocompatibility \cite{Hatakeyama2012, Pissis2013}, no desorption was evidenced. Depending on the water content, typically for a hydration of $\sim$ 20 wt.\% , only crystallisation at sub-zero temperatures was ob!
 served inside the matrix \cite {Pissis2005,Gutierrez2008}.

Another typical example of soft confinement is given by hydrated surfactant. In water-oil reverse micelles of typically 1-5  nanometers core diameter, supercooled water can be observed down to several tens of Kelvin, but the ideal solution behaviour did not justify the large freezing point depressions observed in these systems \cite{Spehr2008}. Nonetheless, surfactants forming lamellar phases separated by interstitial water seem to exhibit ultra viscous water: the water confined in model lipid membranes is also transported in and out of the layers at sub-zero temperatures \cite{Gleeson1994}, forming ice pools in contact with the non-freezing interstitial water. As in the Nafion case, the ice formation gradually increases upon temperature decrease, in a reversible way, with a little hysteresis assigned to water supercooling. The same effect has been observed in the natural Purple Membrane \cite{Weik2005,Lechner1998}, between 250 and 273 K. The kinetics of water migration in a!
 nd out of the interlamellar spaces were evidenced, with timescales similar to those measured in Nafion. 

Extensive studies of water transport in flash-cooled protein crystals have been conducted with the aim of improving the quality of crystallographic data. The crystals are usually quenched to 100 K  and during the annealing phases, around 230-250 K, liquid-like water can be observed to be transported in and out of the crystals\cite {Weik2005b, Kriminski2002, Juers2004}. 

Eventually, montmomillorite, a silicate based clay in which water is confined together with various cations, also exhibits the same phenomena \cite{Anderson1967}, showing that the effect is not restricted to soft condensed matter.

These few examples show that systems in which water flows at low temperature are numerous and diversified, organic and inorganic, molecular or polymeric, always ionic. The water behaviour of desorption or freezing is however still unpredicted. Further studies on model systems containing confined water with different physico-chemical properties would therefore be necessary to clarify the relevant parameters relative to the observation of reversible water desorption below 273 K.

\section{Conclusion}
In this work, we quantitatively characterised the water sorption/desorption process in the Nafion membrane between 180 and 270 K by neutron diffraction. In agreement with other studies, we showed that ice crystallises outside of the membrane, in the hexagonal form. The desorption upon cooling is also reflected, in the diffraction patterns, by the shift towards higher Q of the position of the ionomer peak, indicating a contraction of the membrane. The phenomena is exactly reversible upon heating, except for a small hysteresis due to the supercooling of water inside the matrix. The amount of ice present in the system, at a given temperature, can be modelled without using any adjustable parameter by the relation between the freezing point depression of a solution and its concentration. This points out the entropic origin of the phenomenon, limiting the water absorption below 273 K.

\section*{Acknowledgements}
This work was supported by REGIONE TOSCANA POR-CRO-FSE 2007-2013 by EC COST Action MP0902-COINAPO. We thank the Insitut Laue Langevin for beam time allocation. We would like to thank Andrea Taschin for his suppport on TG experiments. MP thanks Martin Weik and Elisabeth Charlaix for fruitful discussions.







\end{document}